\documentclass[print, twocolumn,showpacs]{revtex4}
\usepackage{amsmath}
\usepackage{txfonts}
\usepackage{graphicx}
\usepackage{bm}
\usepackage{epsfig}
\usepackage{setspace}

\begin{document}
\draft
\bibliographystyle{revtex4}

\titlepage
\title{Quantum correlation in three-qubit Heisenberg model with Dzyaloshinskii-Moriya interaction}
\author{Li-Jun Tian$^{1,2}$}\author{Yi-Ying Yan\footnote{Corresponding author. Email: yanyiying@shu.edu.cn.}$^{1}$}\author{Li-Guo Qin$^{1}$}
\affiliation {$^{1}$Department of Physics, Shanghai University, Shanghai,
200444, China}\affiliation{$^{2}$Shanghai Key Lab for Astrophysics, Shanghai,
200234, China}

\begin{abstract}

    We investigate the pairwise thermal quantum discord in a
    three-qubit XXZ model with Dzyaloshinskii-Moriya (DM) interaction. We find that the DM interaction can increase quantum discord to a fixed value
    in the antiferromagnetic system, but decreases quantum discord to a minimum first, then increases it to a fixed value
    in the ferromagnetic system. Abrupt change of quantum discord is observed, which indicates the abrupt change of groundstate. Dynamics of pairwise thermal quantum discord is also considered. We show that thermal discord vanishes in
    asymptotic limit regardless of its initial values, while thermal entanglement suddenly disappears at finite time.

\end{abstract}
\maketitle

\section{Introduction}
Entanglement has been intensively investigated in recent years as it is
thought to be a fundamental resource for quantum computational
processing tasks, e.g., quantum computation, teleportation and dense
coding \cite{Nielsen1}. It is also considered to be the most
nonclassical nature of quantum composite systems and impossible to
be simulated within classical formalism \cite{Horodecki1}. Recently, it is found that there may be nonclassical correlations in the
quantum states other than entanglement, which is
fragile with respect to environment. Quantum discord (QD) is proposed to be a measure of
quantumness of correlation of bipartite systems, which is arising from
the difference between two quantum extensions of the classical
mutual information \cite{Zurek1}. In the case of pure
states, QD reduces to entropy of entanglement \cite{Luo1}. In mixed states, QD and
entanglement are independent measures of correlation without simple
relative ordering \cite{Luo1, Ali1}. QD of disentangled state may not be zero \cite{Zurek1}. Nonentanglement quantum
discord may provide the quantum advantage in the deterministic quantum computation with one quantum
bit (DQC$1$)\cite{Datta1,Lanyon1}. Though quantum discord arises from quantum information, it may also be applied to condensed matter
physics, such as indicating the quantum phase transition
\cite{Werlang2, Dillenschneider1, Werlang3, Chen1}.

Spin is thought to be a suitable candidate as qubits, and the
Heisenberg model describes the basic spin-spin interaction
\cite{Vala1}. Naturally, this type of model has been extensively
studied, and can be found in Refs.~\cite{Ciliberti1,Maziero1, Hassan1, Werlang2, Pal1}. Some features of QD are revealed, and
behavior of QD is characterized with tunable parameters, e.g., external field,
temperature, spin-spin coupling constant etc. By changing the
temperature and also by applying an external field, several remarkable effects for QD are observed, many of them in sharp contrast to
the behavior observed for the entanglement. It is revealed that situations where QD increases with temperature while entanglement decreases or while entanglement is zero in two-qubit Heisenberg model \cite{Werlang2}. It is argued that models of low-dimensional magnetic materials may supplemented with Dzyaloshinsky-Moriya (DM) interaction, which is arising from the spin-orbit coupling \cite{Moriya1}.
The Dzyaloshinsky-Moriya exchange interaction describes superexchange between the interacting spins, and is known to generate many dramatical features \cite{Maruyama1, Zhang1, Kargarian1, Hao1, Cheng2}. It is found that DM interaction can excite entanglement and teleportation fidelity.

Decoherence is the main obstacle for practical implementation of
quantum computing. Dynamics of quantum discord in presence interaction with the environment can be
found in Refs.~\cite{Werlang1, Fanchini1, Maziero2, Maziero3, Li2, Xiao1}. Both Markovian and non-Markovian dynamics of QD were investigated theoretically and experimentally.
It was revealed that quantum discord is more robust than entanglement
against decoherence. In this sense, it is argued that the quantum
algorithms based only on quantum discord is expected to be more
robust than based on entanglement. Recently, Quantum correlation has been investigated experimentally \cite{Xu1}, the sudden transition from a classical to a quantum decoherence regime
is observed during the dynamics of a Bell diagonal state in a
non-Markovian dephasing environment.

In this paper we consider pairwise QD of three-qubit XXZ model
with Dzyaloshinskii-Moriya interaction in a heat reservoir. We start in Sec. \ref{sec2} for introducing three qubits XXZ model with DM interaction in a canonical ensemble. In Sec.~\ref{sec4} we investigate behavior of pairwise thermal QD with variation of system parameters, and present the respective numerical results. In Sec.~\ref{sec5}, We calculate the dissipative dynamics of pairwise thermal quantum discord under Markovian environments. We analyze two types of quantum channel such as dephasing, depolarizing, by assuming each qubit coupling to local noisy independently. The conclusions are obtained in the last section.

\section{Model and its SOLUTION}\label{sec2}
The Hamiltonian for a XXZ model with $z$-component DM
interaction can be expressed as
\begin{eqnarray}\label{e2.1}
H&=&\frac {J}{2} \sum_{n=1}^{3}[\sigma^{x}_{n} \sigma^{x}_{n+1}+
\sigma^{y}_{n}\sigma^{y}_{n+1}+ \Delta
\sigma^{z}_{n}\sigma^{z}_{n+1} \nonumber \\
& & {}+D(\sigma^{x}_{n}\sigma^{y}_{n+1}-\sigma^{y}_{n}\sigma^{x}_{n+1})]+B\sum_{n=1}^{3}\sigma^{z}_{n},
\end{eqnarray}
where $\sigma^{\alpha}_{n}(\alpha=x,y,z)$ denote the Pauli matrices for qubit $n$,
 $J$ is the coupling constant, $D$ is the
$z$-component DM interaction strength, $\Delta$ is an anisotropy
parameter, $B$ is an external magnetic field in the $z$-direction. Periodic
boundary conditions are assumed as $\sigma^{\alpha}_{1}=\sigma^{\alpha}_{4}$.

It is easy to calculate the eigenstates of the
Hamiltonian:

\begin{eqnarray} \label{e2.2}
|\phi_{0}\rangle&=&|000\rangle, \nonumber \\
|\phi_{1}\rangle&=&\frac{1}{\sqrt{3}}(|011\rangle+|101\rangle+|110\rangle),\nonumber  \\
|\phi_{2}\rangle&=&\frac{1}{\sqrt{3}}(|001\rangle+|010\rangle+|100\rangle),\nonumber
\\
|\phi_{3}\rangle&=&\frac{1}{2\sqrt{3}}[(i+\sqrt{3})|011\rangle+(i-\sqrt{3})|101\rangle-2i|110\rangle],\nonumber
\\
|\phi_{4}\rangle&=&\frac{1}{2\sqrt{3}}[(i-\sqrt{3})|011\rangle+(i+\sqrt{3})|101\rangle-2i|110\rangle],\nonumber
\\
|\phi_{5}\rangle&=&\frac{1}{2\sqrt{3}}[(i+\sqrt{3})|001\rangle+(i-\sqrt{3})|010\rangle-2i|100\rangle],\nonumber
\\
|\phi_{6}\rangle&=&\frac{1}{2\sqrt{3}}[(i-\sqrt{3})|001\rangle+(i+\sqrt{3})|010\rangle-2i|100\rangle],\nonumber
\\
|\phi_{7}\rangle&=&|111\rangle,
\end{eqnarray}
and the corresponding eigenvalues: $E_{0}=\frac{3}{2}J\Delta+3B$, $E_{1}=2J-\frac{1}{2}J\Delta-B$, $E_{2}=2J-\frac{1}{2}J\Delta+B$, $E_{3}=-B-J-\frac{1}{2}J\Delta-\sqrt{3}JD$, $E_{4}=-B-J-\frac{1}{2}J\Delta+\sqrt{3}JD$,
$E_{5}=B-J-\frac{1}{2}J\Delta-\sqrt{3}JD$, $E_{6}=B-J-\frac{1}{2}J\Delta+\sqrt{3}JD$, $E_{7}=\frac{3}{2}J\Delta-3B$, respectively.

For a system in thermal equilibrium, the density matrix $\rho(T)=\exp(-\beta H)/Z$, where $\beta=1/k_{B}T$ ($k_{B}$ is Boltzmann's constant, which we set it equal to $1$) and $Z=Tr[\exp(-\beta H)]$ is the partition function. Hence, in this model, we have
\begin{equation}\label{e2.4}
\rho(T)=\frac{1}{Z}\sum_{i=0}^{7}\exp(-\beta
E_{i})|\phi_{i}\rangle\langle\phi_{i}|,
\end{equation}
with $Z=\sum_{i=0}^{7}\exp(-\beta E_{i})$. The two-qubit reduced density matrix is obtained by tracing all but the first two spins, namely, $\rho_{12}(T)=Tr_{3}[\rho(T)]$. Due to the periodic boundary conditions, it is easy to check that all of the reduced matrices are equal.
In the standard basis
$\{|00\rangle, |01\rangle, |10\rangle, |11\rangle \}$ the density matrix $\rho_{12}(T)$ is given by

\begin{equation}\label{e2.51}
\rho_{12}(T)= \frac{1}{6Z^{\prime}}\left (
\begin{array}{cccc}
u & 0 & 0 & 0 \\
0 & w & y & 0 \\
0 & y^{*} & w & 0 \\
0 & 0 & 0 & v \\
\end{array}
\right )
\end{equation}
where
\begin{eqnarray} \label{e2.6}
u&=&3\exp{(-3\beta B)}+\exp{[\beta (2J\Delta-B)]}[\exp{(-2\beta
J)}\nonumber \\
& &+2\cosh(\sqrt{3}\beta JD)\exp{(\beta J)}],\nonumber
\\
v&=&3\exp{(3\beta B)}+\exp{[\beta (2J\Delta+B)]}[\exp{(-2\beta
J)}\nonumber \\
& &+2\cosh(\sqrt{3}\beta JD)\exp{(\beta J)}],\nonumber
\\
w&=&2\cosh(\beta B)\exp{(2\beta J\Delta)}[\exp{(-2\beta J)}\nonumber \\
& & +2\cosh(\sqrt{3}\beta JD)\exp{(\beta J)}], \nonumber \\
y&=&2\cosh(\beta B)\exp{(2\beta J\Delta)}[\exp{(-2\beta J)}-\nonumber \\
& &\cosh(\sqrt{3}\beta JD)\exp{(\beta J)}-i\sqrt{3}\sinh(\sqrt{3}\beta JD)\exp{(\beta J)}], \nonumber \\
Z^{\prime}&=&\cosh(3\beta B)+\cosh(\beta B)\exp{(2\beta J\Delta)}[\exp{(-2\beta J)} \nonumber \\
& &+2\cosh(\sqrt{3}\beta JD)\exp{(\beta J)}],
\end{eqnarray}
and $y^{*}$ represents the complex conjugation of $y$. Quantum discord of this state is called pairwise thermal quantum discord. In the paper we normalize the coupling constant between spins to $J=-1$ for the ferromagnetic case and $J=1$ for the
antiferromagnetic case and study QD for the two cases
separately.

\section{Thermal quantum discord }\label{sec4}
The quantum discord, which measures the quantum correlation of bipartite system $\rho_{AB}$, is defined as difference of two versions of quantum
mutual information \cite{Zurek1},
\begin{equation}
Q(\rho_{AB})=I(\rho_{AB})-C(\rho_{AB}),
\end{equation}
where
\begin{equation}
I(\rho_{AB})=S(\rho_{A})+S(\rho_{B})-S(\rho_{AB}),
\end{equation}
is the quantum mutual information, which measures the total correlation of system $\rho_{AB}$. $S(\rho)=-Tr(\rho \log \rho)$ is von Neumann entropy.
$\rho_{A(B)}$ is the reduced state of $\rho_{AB}$. And
\begin{equation}
C(\rho_{AB})=\max_{\{\Pi_{B}^{i}\}}\{S(\rho_{A})-\sum_{i}p_{i}S(\rho_{A}^{i})\},
\end{equation}
is the quantum conditional entropy, which is the maximum information of subsystem $A$ obtained by performing a measurement on subsystem $B$ measures classical correlation \cite{Vedral1}. $\{\Pi_{B}^{i}\}$ represents a set of von Neumann measurement on $B$.
$\rho_{A}^{i}=Tr_{B}(\Pi_{B}^{i}\rho_{AB}\Pi_{B}^{i})/Tr_{AB}(\Pi_{B}^{i}\rho_{AB}\Pi_{B}^{i})$ is the state of $A$ after obtaining outcome $i$ on $B$, and $p_{i}= Tr_{AB}(\Pi_{B}^{i}\rho_{AB}\Pi_{B}^{i})$.

To compute QD between spins, we choose the set of projectors $\{\Pi^{i}_{B}=V|i\rangle\langle i|V^{\dag}| i=0,1, V \in SU(2)\}$ for subsystem $B$. Following the procedure of Ref. \cite{Ali1}, we have the QD for the reduced thermal state (\ref{e2.51}) as follow,
\begin{equation} \label{e2.19}
Q(\rho_{12}(T))=S(\rho_{1})+\sum_{k=0}^{3} \lambda_{k} \log_{2}
\lambda_{k}+\min\{S_{1}, S_{2}\},
\end{equation}
where $\rho_{1}$ is the reduced state of $\rho_{12}(T)$, $\lambda_{k}$ is the eigenvalue of the $\rho_{12}(T)$, $S_{1}=p_{0}h(\theta_{0})+p_{1}h(\theta_{1})$ and $S_{2}=h(\theta_{2})$
with $h(\theta)=-\frac{1-\theta}{2}\log_{2}\frac{1-\theta}{2}-\frac{1+\theta}{2}\log_{2}\frac{1+\theta}{2}$,
 the parameters can be expressed in terms of matrix elements as
$\theta_{0}=\left|(u-w)/(u+w)\right|$, $\theta_{1}=\left|(w-v)/(w+v)\right|$,
$\theta_{2}=\sqrt{(u-v)^{2}+4|y|^{2}}/6Z^{\prime}$, $p_{0}=(u+w)/6Z^{\prime}$, and $p_{1}=(w+v)/6Z^{\prime}.$

\begin{figure}
  % Requires \usepackage{graphicx}
  \includegraphics[width=8cm]{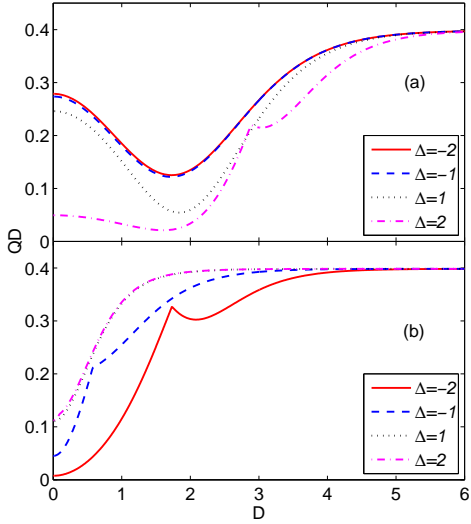}
  \caption{(Color online) thermal quantum discord as a function of $D$ at the temperature $T=0.9$, $B=0$, (a) $J=-1$ and (b) $J=1$}\label{Fig.1}
\end{figure}

It is easy to verify that invariance of QD under $D\leftrightarrow-D$, thus we just discuss the positive values of $D$. In Fig.~\ref{Fig.1}, we plot QD as a function of DM interaction while other parameters are fixed $(T=0.9, B=0)$. In the ferromagnetic case $(J=-1)$, as increasing $D$, QD decreases to a minimum value first and then increases to 0.3984. When $D\rightarrow+\infty$, the density matrix (\ref{e2.4}) is given by $\rho=\frac{1}{2}(|\phi_{4}\rangle \langle\phi_{4}|+|\phi_{6}\rangle \langle\phi_{6}|)$ for any finite temperature, the pairwise QD of this state is $0.3984$. We notice that QD may indicate abrupt change of groundstate. When we set $B=0$, and $\Delta=2$, the groundstate depends on $D$, it may be
\begin{equation}
\begin{cases}
\frac{1}{2}(|\phi_{0}\rangle \langle\phi_{0}|+|\phi_{7}\rangle \langle\phi_{7}|)& \text{$D\in[0,\frac{5}{\sqrt{3}})$},\\
\frac{1}{4}(|\phi_{0}\rangle \langle\phi_{0}|+|\phi_{4}\rangle \langle\phi_{4}|+|\phi_{6}\rangle \langle\phi_{6}|+|\phi_{7}\rangle \langle\phi_{7}|)& \text{$D=\frac{5}{\sqrt{3}}$}, \\
\frac{1}{2}(|\phi_{4}\rangle \langle\phi_{4}|+|\phi_{6}\rangle \langle\phi_{6}|)& \text{$D\in(\frac{5}{\sqrt{3}}, +\infty)$}.
\end{cases}
\end{equation}
As tuning $D$ from $0$ to $6$, QD of the groundstate abruptly changes around the critical point $D=\frac{5}{\sqrt{3}}$ : $0 \rightarrow0.3333 \rightarrow 0.3984$. Raising temperature mixes the groundstate with excited state. Thus QD abruptly changes in the plot nearby the point $D=5/\sqrt{3}\approx2.8868$. In the antiferromagnetic case $(J=1)$, QD increases rapidly as $D$ increasing, then tends to a fixed value. It is easy to check that the density matrix (\ref{e2.4}) is $\frac{1}{2}(|\phi_{3}\rangle \langle\phi_{3}|+|\phi_{5}\rangle \langle\phi_{5}|)$ at any finite temperature when $D\rightarrow+\infty$, and the pairwise QD of this state is also equal to $0.3984$. We observe that QD abruptly changes around two points. One can easily obtain the critical point of dashline, $D=\sqrt{3}$ and solidline, $D=\frac{\sqrt{3}}{3}$, where groundstate abruptly changes. Due to the temperature, they are not the exactly equal to the ones in the Fig.~\ref{Fig.1}(b).

\begin{figure}
  % Requires \usepackage{graphicx}
  \includegraphics[width=8cm]{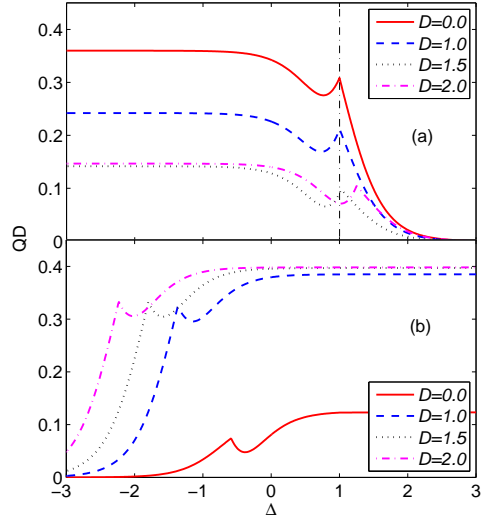}
 \caption{(Color online) thermal quantum discord as a function of $\Delta$ at the temperature $T=0.6$, $B=0$, (a) $J=-1$ and (b) $J=1$}\label{Fig.2}
\end{figure}

We now move to investigate how QD behaves as we change the
anisotropy parameter at finite temperature. In
Fig.~\ref{Fig.2}, we plot how QD depends on anisotropy parameter for
$T=0.6$ and zero magnetic field. Pairwise thermal entanglement of
three-qubit Heisenberg model has been considered in Ref.~
\cite{Wang1}. The authors found that thermal entanglement is absent from
both the antiferromagnetic and the ferromagnetic XXZ
model with anisotropy parameter $\Delta\geq1$. However, there do
exist nonzero values of QD in the vicinity of $\Delta\geq1$. Once again we observe the abrupt change points
in the plots. This behavior has been observed in
$N-$qubit XXZ model \cite{Werlang3}. It is argued that abrupt change of QD in the critical point of $N-$qubit XXZ model is related to the maximization process of QD. The calculation shows that the groundstate abruptly changes at critical point $\Delta=1$ when $D<\sqrt{3}$.
Hence we notice QD abruptly changes on the point $\Delta=1$ for $D<\sqrt{3}$. When
$D=2$, the critical point is $\Delta=\sqrt{3}-0.5\approx1.2321$ near the abrupt change point of dot-dashline. The critical points we obtained may be not the exact point in the plot. This is because the energy gap between groundstate and first excited state is small. As depicted in Fig.~\ref{Fig.2}(a), the abrupt change of dotted line ($D=1.5$) shift from the critical point, this is because first excited state is close to the groundstate, the energy gap is $\Delta E=E_{4}-E_{0}\approx0.4019$.
 Comparing with ferromagnetic case, the
QD of antiferromagnetic case behaves in a less different way as tuning the
anisotropy parameter other than the abrupt change points are located on the
different points. One can readily check that the abrupt change of QD is related to abrupt change of groundstate. Fig.~\ref{Fig.2} shows QD increases to fixed values with $J\Delta$ in both cases. The state of system would be $\frac{1}{2}(|\phi_{0}\rangle\langle\phi_{0}+|\phi_{7}\rangle\langle\phi_{7}|)$ when $J\Delta\rightarrow-\infty$, thus QD vanishes.

\begin{figure}
  % Requires \usepackage{graphicx}
  \includegraphics[width=8cm]{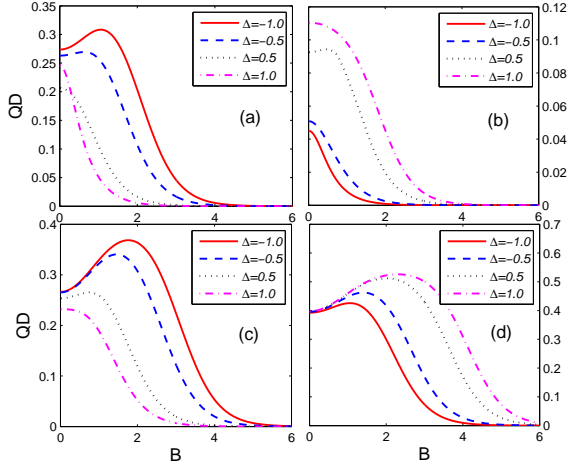}
  \caption{(Color online) thermal quantum discord as a function of $B$ at temperature $T=0.9$, (a) $J=-1$, $D=0$, and (b) $J=1$, $D=0$, and (c) $J=-1,  D=3$, and (d) $J=1, D=3$.}\label{Fig.3}
\end{figure}

We now turn to characterize the dependence of QD on magnetic field $B$. It is easy to check that invariance of QD under magnetic filed $B\leftrightarrow-B$. We just discuss the positive magnetic field $B$. Fig.~\ref{Fig.3}(a)$-$3(b) show the behavior of QD of systems without DM interaction as changing $B$. The QD of ferromagnetic system may increase to a maximum, and then decrease. It is worth noting that magnetic field plays a beneficial role when $J\Delta>0$ in the ferromagnetic system. However, in the antiferromagnetic system, magnetic field just slightly enhances QD when $\Delta=0.5$. The behavior of QD is plotted in Fig.~\ref{Fig.3}(c)-3(d) when DM interaction presents. By applying magnetic field, the value of QD supplemented with DM interaction may have a more larger range than the cases without DM interaction. However, QD finally vanishes as $B$ increases, namely, strong magnetic field is harmful to QD. This is because density matrix (\ref{e2.4}) is $|\phi_{7}\rangle\langle\phi_{7}|$ (separable state) in this limit.

\section{THERMAL STATE UNDER LOCAL DECOHERENCE}\label{sec5}
Now we will focus on decoherence of thermal state when each qubit
couples to local quantum noisy independently. The dynamics of three
qubits interacting independently with individual environments is
described by the solutions of the appropriate Born-Markov-Lindblad
equations \cite{Carmichael1}, which can also be obtained by so
called the Kraus operator approach \cite{Nielsen1}. Given the
initial state $\rho(0)$, its evolution equation can be written as
\begin{equation} \label{e4.24}
\rho(t)=\sum_{\alpha, \beta, \tau} E_{\alpha, \beta,
\tau}\rho(0)E^{\dagger}_{\alpha, \beta, \tau},
\end{equation}
where the Kraus operators, $E_{\alpha, \beta, \tau}=E_{\alpha}\otimes
E_{\beta}\otimes E_{\tau}$ \cite{Nielsen1} satisfy the
completeness condition $\sum_{\alpha, \beta, \tau} E_{\alpha, \beta,
\tau}^{\dagger}E_{\alpha, \beta, \tau}=I$. The operators
$\{E_{\alpha}\}$ describe the quantum channel effects of one qubit.

\begin{figure}
  % Requires \usepackage{graphicx}
  \includegraphics[width=8cm]{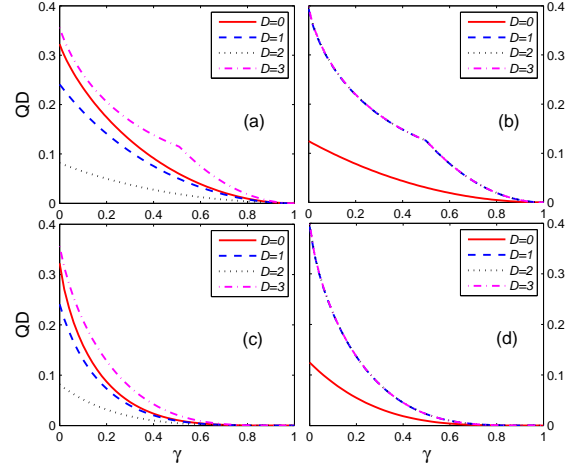}
  \caption{Dissipative dynamics of QD as a function of $\gamma$ at the temperature $T=0.5$, $B=0$, $\Delta=1$, (a) $J=-1$ and (b) $J=1$ for dephasing channel, and (c) $J=-1$, and (d) $J=1$ for depolarizing channel.}\label{Fig.4}
\end{figure}

The dephasing channel, which describes the loss of quantum information without any energy
dissipation \cite{Nielsen1}, is given by
\begin{eqnarray} \label{e4.25}
E_{0}&=& \left (
\begin{array}{cc}
1 & 0 \\
0 & \sqrt{1-\gamma} \\
\end{array}
\right ), \nonumber \\
E_{1}&=& \left (
\begin{array}{cc}
0 & 0 \\
0 & \sqrt{\gamma} \\
\end{array}
\right ),
\end{eqnarray}
where $\gamma=1-e^{-\chi t}$ and $\chi$ denotes the decay rate
associating with real physical process. To study pairwise QD, we trace
out the $3rd$ site. The reduced density matrix of any pair of sites are equal because of the symmetry of the Hamiltonian
and assumption of qubits coupling to environments independently.
Taking the Eq.~(\ref{e2.4}) as the initial state and using Eqs.~(\ref{e4.24}) and (\ref{e4.25}), we have time-dependent state
\begin{equation}\label{e4.26}
\rho_{12}(t)|_{T}= \frac{1}{6Z^{\prime}}\left (
\begin{array}{cccc}
u & 0 & 0 & 0 \\
0 & \emph{w} & (1-\gamma)y & 0 \\
0 &(1-\gamma)y^{*} & w & 0 \\
0 & 0 & 0 & v \\
\end{array}
\right ),
\end{equation}
where $\gamma$ and other parameters were defined above. We investigate the behavior of QD and entanglement in this state without magnetic filed as follow. Substituting Eq.~(13) into Eq.~(9), we can obtain the QD \cite{Luo1, Ali1},
\begin{eqnarray} \label{e4.29}
Q&=&-\frac{u_{0}+w_{0}}{3Z^{\prime}_{0}}\log_{2}\frac{u_{0}+w_{0}}{6Z^{\prime}_{0}}+\frac{u_{0}}{3Z^{\prime}_{0}}\log_{2}\frac{u_{0}}{6Z^{\prime}_{0}}
\nonumber \\
& & +\sum_{l=\pm1}\bigg[\frac{w_{0}+l(1-\gamma)|y_{0}|}{6Z^{\prime}_{0}}\log_{2}\frac{w_{0}+l(1-\gamma)|y_{0}|}{6Z^{\prime}_{0}} \nonumber
\\ & & -\frac{1+l\theta}{2}\log_{2}\frac{1+l\theta}{2}\bigg],
\end{eqnarray}
where $u_{0}$, $w_{0}$, $v_{0}$, $y_{0}$ and $Z^{\prime}_{0}$ represent the
parameters defined in Eq.~(\ref{e2.6}) for $B=0$ respectively, and $\theta=\max\{|u_{0}-w_{0}|/3Z_{0}^{\prime}, (1-\gamma)|y_{0}|/3Z_{0}^{\prime}\}$. To compare the behavior of entanglement with QD, we investigate the pairwise entanglement by
using concurrence as the quantifier \cite{Wootters1}. One can
write out the concurrence ($C$) for Eq.~(\ref{e4.26}) \cite{Wang1},
\begin{equation} \label{e4.27}
C=\frac{1}{3Z^{\prime}_{0}}\max\{0,
(1-\gamma)|y_{0}|-\sqrt{u_{0}v_{0}}\}.
\end{equation}

\begin{figure}
  % Requires \usepackage{graphicx}
  \includegraphics[width=8cm]{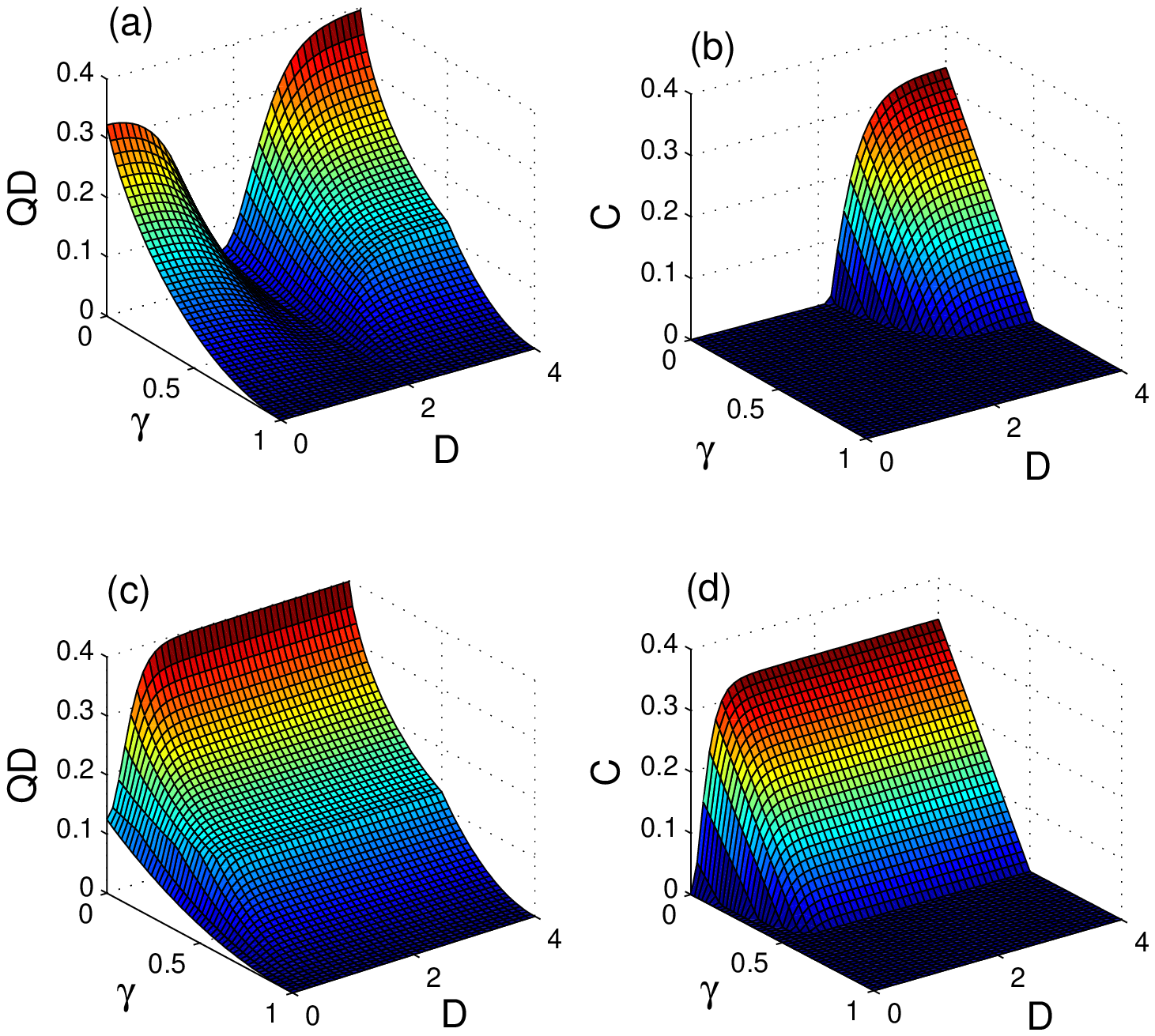}
  \caption{Dissipative dynamics of QD and concurrence as a function of $\gamma$ and $D$, at the temperature $T=0.5$, $B=0$, $\Delta=0.5$. (a) and (c) for QD; (b) and (d) for concurrence. (a) and (b) $J=-1$; (c) and (d) $J=1$}\label{Fig.5}
\end{figure}

Figure.~\ref{Fig.4}(a)$-$4(b) show the numerical result for qubits coupling to dephasing channel. We notice that the derivative, $dQ/d\gamma$, of dot-dash line $(D=3)$ in Fig.~\ref{Fig.4}(a) abruptly changes around a point. The similar behavior is observed in the antiferromagnetic system with nonzero $D$. The QD regardless of its initial value only disappear in the limit $\gamma\rightarrow1$. Fig.~\ref{Fig.5}(a)$-$5(d) shows the behavior of quantum discord and entanglement with variation of $D$ and $\gamma$ for fixed temperature $T=0.5$ and $\Delta=0.5$. QD decreases to zero when $\gamma=1$ ($t\rightarrow \infty$) in both cases of $J$.
Obviously, entanglement ($C$) decreases linearly with $\gamma$ for fixed $D$ in the plot Fig.~\ref{Fig.5}(b)$-$5(d), it undergoes a sudden death regardless of $J$ and initial value of $C$ at finite
time.

\begin{figure}
  % Requires \usepackage{graphicx}
  \includegraphics[width=8cm]{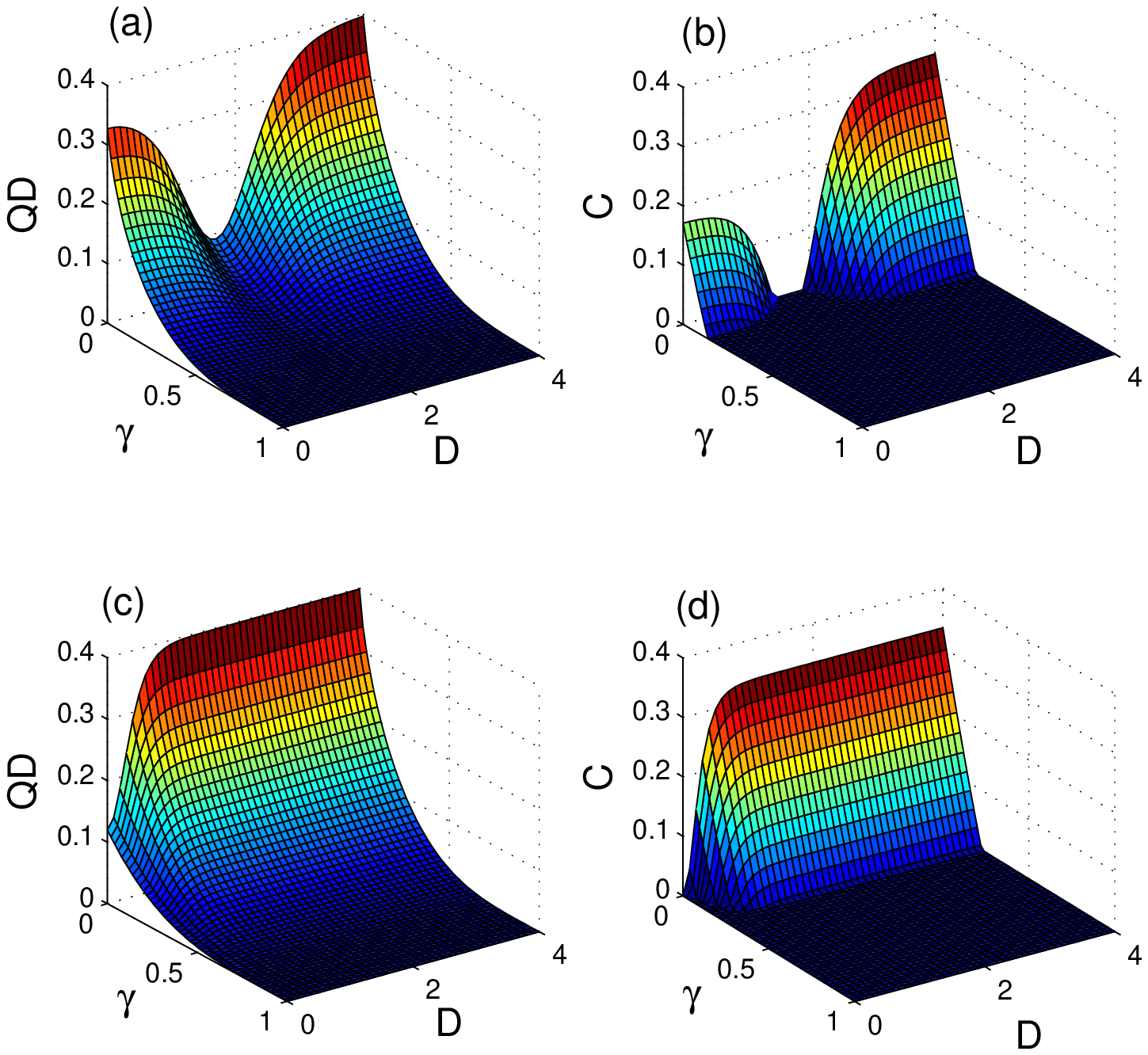}
  \caption{Dissipative dynamics of QD and concurrence as a function of $\gamma$ and $D$, at the temperature $T=0.5$, $B=0$, $\Delta=0.5$. (a) and (c) for QD; (b) and (d) for concurrence. (a) and (b) $J=-1$; (c) and (d) $J=1$}\label{Fig.7}
\end{figure}

The depolarizing channel is an important type of quantum noise, it
describes the process in which the density matrix is dynamically
replaced by the state $I/2$, $I$ denoting identity matrix of a
qubit. The Kraus operators simulate the effect of the depolarizing
channel is given by \cite{Nielsen1}
\begin{eqnarray} \label{e4.30}
E_{0}&=& \sqrt{1-3\gamma/4}\left (
\begin{array}{cc}
1 & 0 \\
0 & 1 \\
\end{array}
\right ), \nonumber \\
E_{1}&=& \sqrt{\gamma/4}\left (
\begin{array}{cc}
0 & 1 \\
1 & 0 \\
\end{array}
\right ), \nonumber \\
E_{2}&=& \sqrt{\gamma/4}\left (
\begin{array}{cc}
0 & -i \\
i & 0 \\
\end{array}
\right ), \nonumber \\
E_{3}&=& \sqrt{\gamma/4}\left (
\begin{array}{cc}
1 & 0 \\
0 & -1 \\
\end{array}
\right ),
\end{eqnarray}
where $\gamma$ was defined above. We use the same procedure as above
to evaluate QD and entanglement numerically. As shown in Fig.~\ref{Fig.4}(c)$-$4(d), QD decreases more fast under this type of noisy than dephasing channel. The derivative, $dQ/d\gamma$, is continuous when qubits couple to this quantum channel, which is different from the case that qubit couples to dephasing channel.
In Fig.~\ref{Fig.7}(a)$-$6(d) we plot the quantum discord and entanglement for the variation of $D$ and $\gamma$. When system
couples to this kind of quantum noisy, entanglement once again
disappears at finite time but QD does not.

\section {Conclusions}
In summary, We investigated the properties of thermal quantum
discord within three-qubit Heisenberg model supplemented with DM interaction. We
found that the DM interaction can both decrease and increase quantum
correlation in the ferromagnetic XXZ system, but just increase quantum
correlation rapidly in the antiferromagnetic XXZ system. The
abrupt change of QD was observed as tuning
parameters such as $D$ and $\Delta$, which implies QD may signal the
abrupt change of groundstate. The effects
of magnetic fields on quantum discord were also considered. Our result shows strong magnetic suppresses quantum correlation, while weak magnetic can increase or decrease QD by controlling $J\Delta$. We utilized thermal state as initial
condition to calculate the dynamics of pairwise QD under Markovian
environments. We found pairwise entanglement of thermal state may
occur sudden death but QD is robust enough to disappear in
asymptotic time.

\begin{acknowledgments}
This work was partially supported by the NSF of China (Grant No.
11075101), and Shanghai Research Foundation (Grant No. 07dz22020).
\end{acknowledgments}

\end{document}